\documentclass[12pt]{article}
\usepackage{latexsym}
\usepackage{amsmath}
\usepackage{amsfonts}
\usepackage{amssymb}
\usepackage{graphicx}

\begin{document}
\input epsf
\def\be{\begin{equation}}
\def\bea{\begin{eqnarray}}
\def\ee{\end{equation}}
\def\eea{\end{eqnarray}}
\def\d{\partial}

\long\def\symbolfootnote[#1]#2{\begingroup%
\def\thefootnote{\fnsymbol{footnote}}\footnote[#1]{#2}\endgroup}
\renewcommand{\a}{\left( 1- \frac{2M}{r} \right)}
\newcommand{\dm}{\begin{displaymath}}
\newcommand{\edm}{\end{displaymath}}
\newcommand{\com}[2]{\ensuremath{\left[ #1,#2\right]}}
\newcommand{\la}{\lambda}
\newcommand{\eps}{\ensuremath{\epsilon}}
\newcommand{\half}{\frac{1}{2}}
\newcommand{\field}[1]{\ensuremath{\mathbb{#1}}}
\renewcommand{\l}{\ell}
\newcommand{\bl}{\left(\l\,\right)}
\newcommand{\normljk}{\langle\l,j,k|\l,j,k\rangle}
\newcommand{\N}{\mathcal{N}}
\renewcommand{\b}[1]{\mathbf{#1}}
\renewcommand{\v}{\xi}
\newcommand{\tr}{\tilde{r}}
\newcommand{\ttheta}{\tilde{\theta}}
\newcommand{\bg}{\bar{g}}

\renewcommand{\implies}{\Rightarrow}
\newcommand{\z}{\ensuremath{\ell_{0}}}
\newcommand{\temp}{\ensuremath{\sqrt{\frac{2\z+1}{\z}}}}
\newcommand{\twomatrix}[4]{\ensuremath{\left(\begin{array}{cc} #1 & #2
\\ #3 & #4 \end{array}\right) }}
\newcommand{\columnvec}[2]{\ensuremath{\left(\begin{array}{c} #1 \\ #2
\end{array}\right) }}
\newcommand{\e}{\mbox{\textbf{e}}}
\newcommand{\gm}{\Gamma}
\newcommand{\bt}{\bar{t}}

\newcommand{\bphi}{\bar{\phi}}

\newcommand{\m}{\ensuremath{\mathbf{m}}}
\newcommand{\n}{\ensuremath{\mathbf{n}}}
\renewcommand{\theequation}{\arabic{section}.\arabic{equation}}

\begin{flushright} OHSTPY-HEP-T-04-005\\ hep-th/0405017
\end{flushright}
\vspace{20mm}
\begin{center} {\LARGE Dual  geometries for a set of 3-charge
microstates}
\\
\vspace{20mm} {\bf  Stefano Giusto, Samir D. Mathur and Ashish Saxena}\\ 
\symbolfootnote[0]{giusto@mps.ohio-state.edu, mathur@mps.ohio-state.edu, ashish@pacific.mps.ohio-state.edu} 
\vspace{4mm} Department of Physics,\\ The Ohio State
University,\\ Columbus, OH 43210, USA\\
\vspace{4mm}
\end{center}
\vspace{10mm}
\begin{abstract}

We construct a set of extremal D1-D5-P solutions, by taking appropriate limits in a known family of nonextremal 3-charge solutions. The extremal geometries turn out to be completely smooth, with no horizon and no singularity. The solutions have the right charges to be the duals of a family of CFT microstates which are obtained by spectral flow from the NS vacuum.

\end{abstract}
\thispagestyle{empty}
\newpage
\setcounter{page}{1}
\section{Introduction}\setcounter{equation}{0}

In the traditional picture of a black hole, infalling matter settles
into a central singularity while Hawking radiation emerges at the
horizon. Due to the large separation between the horizon and the
singularity the radiation is insensitive to the detailed state of the
matter that made the hole, and we get information loss \cite{hawking}.

Some computations in string theory suggest that the black hole
interior is quite different; instead of `empty space with a central
singularity' we have a `fuzzball' with state information
distributed throughout the interior of the horizon.  It was shown
in \cite{emission} that due to the phenomenon of `fractionation' the
effective excitations of a D1-D5-P bound state are very light, and
in fact extend to a distance of order the horizon radius.

In
\cite{lm4,lm5} the 2-charge extremal D1-D5 system was studied.
The `naive' geometry of D1 and D5 branes is pictured in Fig.1(a); we
have flat space at $r\rightarrow \infty$ and a singularity at $r=0$.
But the CFT analysis implies that the Ramond (R) ground state of
the D1-D5 system is highly degenerate, with entropy
$S=2\sqrt{2}\sqrt{n_1n_5}$.  In \cite{lm4} the geometries dual to
these states were constructed.  It was found that the naive metric
 did not arise from any of the microstates; instead all
states yielded geometries that were `capped' smoothly before
reaching $r=0$.\footnote{The construction in \cite{lm4} had an
apparent singularity along a closed curve in the `cap', but it was
shown in \cite{lmm} that this was just a coordinate singularity.}
No individual geometry has a horizon or  singularity but if we
  draw a surface to bound the area where these geometries
differ significantly from the naive geometry then from
the area $A$ of this surface we find
\be
{A\over 4G} \sim 2\sqrt{2}\sqrt{n_1n_5}
\label{two}
\ee
The radius of this surface is $\sim ({n_1n_5})^{1\over 6}$ times the
Planck length or the string length (the dilaton is bounded, so
$l_p\sim l_s$). Thus we see that the D1-D5 bound state `swells' up to a
  radius that increases with the charges,  and which is such that
the bounding surface constructed above  bears  a Bekenstein
type relation to the count of states.

If we have three charges  --  D1, D5  and momentum P  -- then the
`naive' geometry is an extreme Reissner-Nordstrom type black hole.
This geometry has a horizon at $r=0$, and continues to a region
$r<0$ which contains a singularity (Fig.1(c)).  The area of the horizon
gives
\be
{A\over 4G}=2\pi\sqrt{n_1n_5n_p}
\label{three}
\ee
  and this exactly
equals the microscopic entropy obtained from a count of D1-D5-P
ground states \cite{stromvafa}. But based on the results above we
are led to ask if the individual states are described by geometries
that `cap off'  before reaching $r=0$ as in Fig.1(d).  For three
charges the radius of the `throat' asymptotes to a constant as we
go down the throat, so the area $A$ obtained at the dashed line in
Fig.1(d) will give (\ref{three}).  Thus the nontrivial question in this
case is whether the geometries dual to 3-charge microstates are
like Fig.1(c) (with a horizon and singularity inside the horizon) or
whether
some effects destroy this naive expectation before we reach $r=0$.
Note that in the 3-charge case  (unlike the 2-charge case) we do
not expect the generic state to be well-described by a classical
geometry; quantum fluctuations can be large. But there would still
be special cases that are in fact well described by a classical metric,
and we can gain insight by constructing these explicitly.

In \cite{mss} a perturbation was constructed on an
extremal 2-charge D1-D5 state that added one unit of P charge.
The equation for linear perturbations was solved to give a regular,
normalizable excitation in the limits of small
$r$ and large $r$, and the solutions were shown to agree to
several orders in the region of overlap. This indicated that at least
this particular 3-charge state was smoothly `capped' as in Fig.1(d), and did
not have a horizon or singularity like Fig.1(c).

In the present paper we obtain {\it exact} geometries dual to a
set of D1-D5-P microstates. These geometries will again turn out
to be capped as in Fig.1(d).  The microstates are not generic
3-charge states; in particular they have a significant amount of rotation.
But the construction does support the general conjecture that all
configurations must suffer modifications before reaching $r=0$ and
forming  a horizon.

\begin{figure}[ht]
\hspace{0in}
\includegraphics{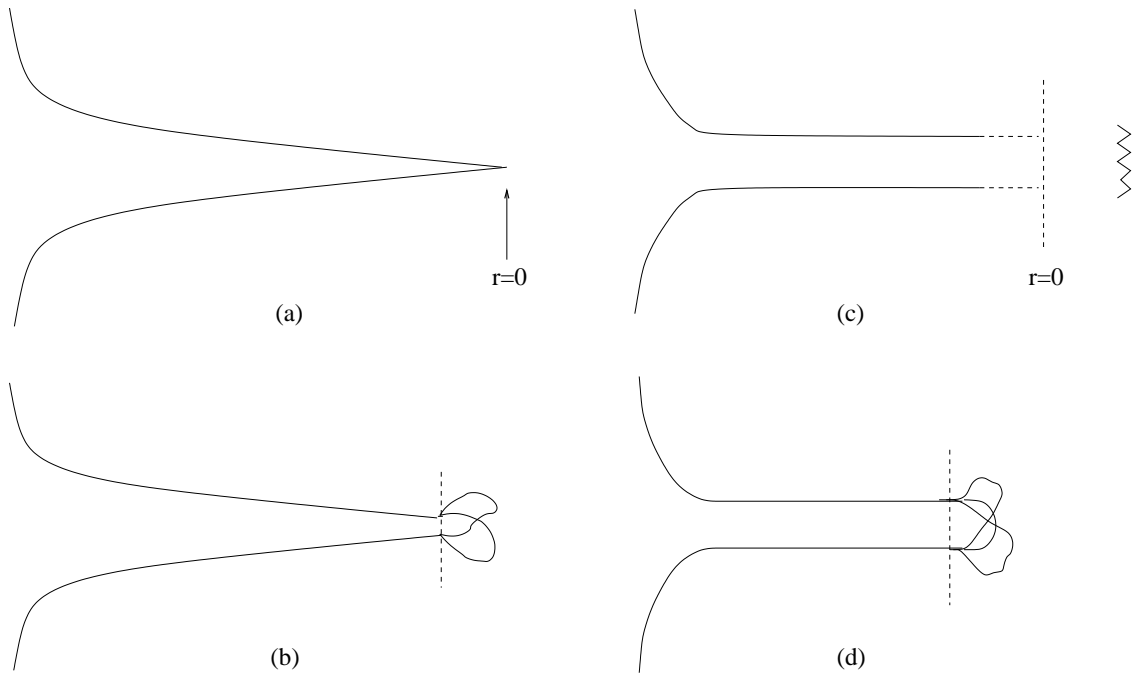}
\caption{ \small{(a) Naive geometry of 2-charge D1-D5. (b) Actual geometries of
2-charge D1-D5; the area of the surface shown by the dashed line gives ${A\over 4G}\sim \sqrt{n_1n_5}$. (c) Naive geometry of 3-charge D1-D5-P; there is a horizon at $r=0$ and a singularity past the horizon.
(d) Expected geometries for D1-D5-P; the area at the dashed line will give ${A\over 4G}=2\pi\sqrt{n_1n_5n_p}$.}} 
\label{fig}
\end{figure}

While we were finishing this work the paper \cite{lunin} appeared,
which also constructed similar metrics by an interesting though different
method based on \cite{gmr}. 
If we set the D1 and D5 charges equal in our solution
($Q_1=Q_5$) then the dilaton vanishes, and we obtain solutions
that look (locally) like the solutions in \cite{lunin}. There does
appear to be a difference however  in the way the final
parameters are set in the solution, so that the values of the
conserved quantities like angular momenta in \cite{lunin} appear to
be different
from the ones that we have. We  comment briefly on these issues
near the end of our paper.

\section{The CFT states}\setcounter{equation}{0}

\subsection{The D1-D5 CFT}

We take IIB string theory compactified to $M_{4,1}\times S^1\times
T^4$.  Let $y$ be the coordinate along $S^1$ with
\be
0\le y<2\pi R
\ee
The $T^4$ is described by 4 coordinates $z_1, z_2, z_3, z_4$, and
the noncompact space is spanned by $t, x_1, x_2, x_3, x_4$.  We
wrap
$n_1$ D1 branes on $S^1$, and
$n_5$ D5 branes on
$S^1\times T^4$. Let $N=n_1n_5$. The bound state of these branes
is described by a 1+1 dimensional sigma model, with base space
$(y,t)$ and target space a deformation of the orbifold
$(T^4)^N/S_N$ (the symmetric product of $N$ copies of $T^4$).  The
CFT has ${\cal N}=4$ supersymmetry, and a moduli space which
preserves this supersymmetry. It is conjectured that in this
moduli space we have an `orbifold point' where the target space is
just the orbifold
$(T^4)^N/S_N$ \cite{sw}.

The CFT with target space just one copy of $T^4$ is described, at the 
orbifold point, by 4
real bosons
$X^1$,  $X^2$,  $X^3$,  $X^4$ (which arise from the 4 directions $z_1, z_2,
z_3, z_4$), 4 real left moving fermions $\psi^1, \psi^2, \psi^3,
\psi^4$ and 4 real right moving fermions $\bar\psi^1, \bar\psi^2,
\bar\psi^3, \bar\psi^4$. The central charge is $c=6$.  The complete
theory with target space $(T^4)^N/S_N$ has $N$ copies of this
$c=6$ CFT, with states that are symmetrized between the $N$
copies. The orbifolding also generates `twist' sectors, which are
created by twist operators $\sigma_n$; for a detailed construction
of the $\sigma_n$ in this theory see \cite{lm1, lm2}. We will not be
working with the twist sectors in this paper -- all the states we
construct are in the `untwisted sector'.

The rotational symmetry of the noncompact directions $x_1\dots
x_4$ gives a symmetry
$so(4)\approx su(2)_L\times su(2)_R$. The left fermions
$\psi^i$ carry spin
${1\over 2}$ under   $su(2)_L$ and the right fermions
$\bar\psi^i$ carry  spin
${1\over 2}$ under   $su(2)_R$.  The `charge' of a state is
given by the quantum numbers $(j, \bar j)=(j^3_L, j^3_R)$.

Consider one copy of the $c=6$ CFT,  and look at the left sector.
The fermions can be antiperiodic around $y$ (NS sector) or
periodic (R sector). The NS vacuum has $h=j=0$. The NS sector
states can be mapped to R sector states by `spectral flow'
\cite{spectral}, under which the conformal dimensions and
charges change as
\bea
h'&=&h-\alpha q + \alpha^2{c\over 24}\\
q'&=&q-\alpha{c\over 12}
\label{qone}
\eea
Setting $\alpha=1$ gives the flow from the NS sector to the R
sector, and we can see that under this flow chiral primaries
of the NS sector
(which have $h=q$) map to Ramond ground states with $h={c\over
24}$.

The field theory on the D1-D5 branes system is in the R sector.
This follows from the fact that the branes are solitons of
the gravity theory, and the fermions on the branes are induced
from fermions on the bulk. The latter are periodic around the
$S^1$; choosing antiperiodic boundary conditions would give a
nonvanishing vacuum energy and disallow the flat space solution
that we have assumed at infinity. The geometries constructed in
\cite{lm4} described gravity duals of the  R ground states of the
CFT.

If we set $\alpha=2$ in (\ref{qone}) then we return to the NS
sector, and setting $\alpha=3$ brings us again to the R sector.
More generally, the choice
\be
\alpha=2n+1, ~~~ n ~~integer
\label{qtwo}
\ee
brings us to the R sector.  From (\ref{qone}) we see that if we start
with a R ground state obtained at $\alpha=1$, then we get another
R ground state ($h={c\over 24}$) at $\alpha=-1$, but for
$\alpha=3,5,7\dots$ and $\alpha=-3,-5, -7\dots$ we get excited
states of the R sector.

\subsection{The states we consider}

We will look at states where we do a spectral flow (\ref{qtwo}) on the
left sector, and a spectral flow with $\alpha=1$ on the right sector.
Thus the right movers will be in  an R ground state, and we get a
supersymmetric configuration of the CFT.
Taking into account all $N=n_1n_5$ copies of the CFT we find
that the states will have dimensions and charges
\bea
h_{total}&=&{1\over 4}(2n+1)^2~n_1n_5\\
j_{total}&=&-{1\over 2} (2n+1)n_1n_5\\
\bar h_{total}&=&{1\over 4} n_1n_5\\
\bar j_{total}&=&-{1\over 2}n_1n_5
\eea
In particular we have `momentum' along the $S^1$, with
\be
h-\bar h=n(n+1) n_1n_5
\label{qfour}
\ee

\subsection{Explicit representation of the states}

Let us construct explicitly the above CFT states. Consider one copy
of the $c=6$ CFT, in the R sector.  The  fermions have modes
$\psi^i_m$. The 4 real fermions can be
grouped into 2 complex fermions  $\psi^+, \psi^-$ which form  a
representation of $su(2)$.  ($\psi^+$ has $j={1\over 2}$ and
$\psi^-$ has $j=-{1\over 2}$.)  The anti-commutation relations
are
\be
\{(\psi^+)^*_m, \psi^+_p\}=\delta_{m+p,0}, ~~~\{(\psi^-)^*_m,
\psi^-_p\}=\delta_{m+p,0}
\ee
  The
$su(2)$ currents are
\be
J^+_m=(\psi^-)^*_{m-p}\psi^+_p,
~~~J^-_m=(\psi^+)^*_{m-p}\psi^-_p, ~~~J^3_m={1\over 2}
[(\psi^-)^*_{m-p}\psi^-_p-(\psi^+)^*_{m-p}\psi^+_p]
\ee
The R ground state for $n=1$ in (\ref{qtwo}) has $j=-{1\over 2}$; we call this
state
$|1\rangle$.
We can get the R ground state with $n=-1$ (which has $j={1\over
2}$) by applying the zero modes $\psi^+_0, (\psi^-)^*_0$, which is
equivalent to applying $J^+_0$
\be
|-1\rangle=
(\psi^-)^*_0\psi^+_0|1\rangle=J^+_0|1\rangle
\ee
We see that we cannot apply $J^+_{-1}$ to $|-1\rangle$, but we can
apply $J^+_{-2}$ getting the state with $n=-2$
\be
|-2\rangle=J^+_{-2}J^+_0|1\rangle=(\psi^-)^*_{-1}\psi^+_{-1}
(\psi^-)^*_0\psi^+_0|1\rangle
\ee
and so on. Returning to the full theory with $n_1n_5$ copies of the
$c=6$ CFT we find that the currents are the sum of the currents in
the individual copies
\be
J^{a,total}_{ n}=(J^a_{n})_1+\dots (J^a_{n})_{n_1n_5}
\ee
and for $k\ge 0$
\be
|-k\rangle^{total}=(J^{+,total}_{-(2k-2)})^{n_1n_5}(J^{+,total}_{-(2k-
4)})^{n_1n_5}
\dots (J^{+,total}_{-2})^{n_1n_5}(J^{+,total}_{0})^{n_1n_5}~|1\rangle^{total}
\ee
Similarly, for $k>1$
\be
|k\rangle^{total}=(J^{-,total}_{-(2k-2)})^{n_1n_5}(J^{-,total}_{-(2k-4
)})^{n_1n_5}
\dots (J^{-,total}_{-2})^{n_1n_5}~|1\rangle^{total}
\ee

\section{Constructing the gravity duals}\setcounter{equation}{0}

In \cite{lm4} the 2-charge D1-D5 solutions were found by dualizing to
the FP system, which has a fundamental
string (F) wrapped on $S^1$ carrying momentum (P) along $S^1$.
Metrics for the vibrating string were constructed,
and dualized back to get D1-D5 geometries. The general geometry was
thus parametrized by the vibration profile
$\vec F(v)$ of the F string. But a 1-parameter subfamily of these
D1-D5 geometries had been found earlier \cite{bal,
mm}, by looking at extremal limits of the general axially symmetric
D1-D5 geometry found in \cite{cy}.

We do not have an analogue of the procedure of \cite{lm4} for
3-charge systems. We will follow instead the
analogue  of \cite{bal, mm} and take an extremal limit of the general
3-charge solution to obtain solutions with D1, D5
and P charges. Taking the limit needs some care, and it will be
important to know in advance the properties of the CFT
states for which we will be finding the duals. The procedure will
give us the duals of the states $|n\rangle^{total}$ which were discussed
in the last section. We will find that the dual geometries are completely
smooth, with no horizon and no singularity.

\subsection{Spectral flow in the gravity description}

In \cite{bal, mm} the following 2-charge D1-D5 solution was found (setting $Q_1=Q_5=Q$ for simplicity)
\bea
ds^2&=&-{1\over h}(dt^2-dy^2)+hf\left( d\theta^2+{dr^2\over
r^2+a^2}\right)-{2aQ\over hf}(\cos^2
\theta dy d\psi+\sin^2\theta dt d\phi)\nonumber\\
&+&h\left[\left(r^2+{a^2Q^2\cos^2\theta\over h^2f^2}\right)\cos^2\theta
d\psi^2+\left(r^2+a^2-{a^2Q^2\sin^2\theta\over h^2f^2}\right)\sin^2\theta
d\phi^2\right]+dz_i dz_i\nonumber\\
\label{el}
\eea
where
\be
a={Q\over R}, ~~~f=r^2+a^2\cos^2\theta, ~~~h=1+{Q\over f}
\ee
Let $R>>\sqrt{Q}$. In the region $r<<\sqrt{Q}$
the geometry (\ref{el}) becomes
\bea
ds^2&=&-{(r^2+a^2\cos^2\theta)\over
Q}(dt^2-dy^2)+Q\left(d\theta^2+{dr^2\over r^2+a^2}\right)\nonumber\\
&&~~~
-2a(\cos^2\theta dyd\psi+\sin^2\theta dtd\phi)+Q(\cos^2\theta
d\psi^2+\sin^2\theta d\phi^2)
\label{innerr}
\eea
The change of coordinates
\be
\psi_{NS}=\psi-\tilde\alpha{a\over Q}y+(\tilde\alpha-1){a\over Q }t, ~~~\phi_{NS}=\phi+(\tilde\alpha-1){a\over Q} y-\tilde\alpha{a\over Q}t
\label{spectral}
\ee
with $\tilde\alpha=1$ brings (\ref{innerr}) to the form $AdS_3\times S^3$
\be
ds^2=-{(r^2+a^2)\over Q}dt^2+{r^2\over Q}dy^2+Q{dr^2\over
r^2+a^2}+Q(d\theta^2+\cos^2\theta d\psi_{NS}^2+\sin^2\theta
d\phi_{NS}^2)
\label{innerns}
\ee
The solution (\ref{el}) describes the asymptotically flat solution created by D1-D5 branes,
with fermions periodic around the $y$ circle. The `near-region' part (\ref{innerr}) describes
the gravity dual of the appropriate CFT state, in the R sector. The change of coordinates
(\ref{spectral}) gives `spectral flow' in the gravity description, and with $\tilde\alpha=1$ we reach (\ref{innerns})
which gives the gravity dual of the NS sector version of the CFT state \cite{bal, mm}.

The 3-charge D1-D5-P states that we wish to describe are obtained by further spectral flow applied to  the state
described by (\ref{innerr}). From (\ref{spectral}) we see that the `near region' geometry dual to these states
is just $AdS_3\times S^3$ described in new coordinates. (The state $|n\rangle^{total}$ is generated by $\tilde\alpha=-n$.)
Thus we may anticipate a `capped' geometry for these states. 
It is not obvious though that the near region solution can be continued to flat space at infinity, while keeping the solution
BPS as well as smooth  (the analogue of (\ref{el})), but we will see that in fact we do get such a solution.

\subsection{The nonextremal 3-charge solution, with angular momentum}

The nonextremal 3-charge metric with rotation was given in \cite{cy},
but we will also need the 2-form gauge field
(which was not listed in \cite{cy}) so we derive this solution in the
Appendix. We follow a different method however
from \cite{cy}; we take a neutral rotating hole, and by a sequence of
boosts and dualities, add the three charges. We
find

\bea
ds^2 &=& - \left( 1- \frac{M\cosh^2\delta_{p}}{f}
\right)\frac{dt^2}{\sqrt{H_{1}H_{5}}}  + \left(1+ \frac{M
\sinh^2\delta_{p}}{f}\right) \frac{ dy^2}{\sqrt{H_{1}H_{5}}}-\frac{M
\sinh2\delta_{p}}{f\sqrt{H_{1}H_{5}}} dt dy \nonumber \\
&+&    f\sqrt{H_{1}H_{5}}\left(\frac{ r^2  dr^2
}{(r^2+a_{1}^2)(r^2+a_{2}^2)-M r^2} +  d\theta^2\right) \nonumber \\
  &+& \left[ (r^2+a_{1}^2) \sqrt{H_{1}H_{5}} + \frac{(a_{2}^2-a_{1}^2)
K_{1}K_{5} \cos^2\theta}{\sqrt{H_{1}H_{5}}} \right] \cos^2\theta
d\psi^2 \nonumber \\
  &+& \left[ (r^2+a_{2}^2) \sqrt{H_{1}H_{5}} +
\frac{(a_{1}^2-a_{2}^2)  K_{1}K_{5} \sin^2\theta}{\sqrt{H_{1}H_{5}}}
\right] \sin^2\theta d\phi^2 \nonumber \\
&+& \frac{M  }{f\sqrt{H_{1}H_{5}}} (a_{1}\cos^2\theta
d\psi+a_{2}\sin^2\theta d\phi)^{2} \nonumber \\
&+& \left.\frac{2 M \cos^2\theta}{f\sqrt{H_{1}H_{5}} }\right[
\left(a_{1}\cosh\delta_{1}\cosh\delta_{5}\cosh\delta_{p} - a_{2}
\sinh\delta_{1}\sinh\delta_{5}\sinh\delta_{p}\right)  dt   \nonumber
\\
& & +  (a_{2}\sinh\delta_{1}\sinh\delta_{5}\cosh\delta_{p}-
a_{1}\cosh\delta_{1}\cosh\delta_{5}\sinh\delta_{p})dy\mbox{\huge
]}d\psi \nonumber \\
&+& \left. \frac{2 M \sin^2\theta}{f\sqrt{H_{1}H_{5}} }\right[
\left(a_{2}\cosh\delta_{1}\cosh\delta_{5}\cosh\delta_{p} - a_{1}
\sinh\delta_{1}\sinh\delta_{5}\sinh\delta_{p}\right)  dt   \nonumber
\\
& &  + (a_{1}\sinh\delta_{1}\sinh\delta_{5}\cosh\delta_{p}-
a_{2}\cosh\delta_{1}\cosh\delta_{5}\sinh\delta_{p})dy\mbox{\huge
]}d\phi + \sqrt{\frac{H_{1}}{H_{5}}}\sum_{i=1}^{4} dz_{i}^2 \nonumber \\
C_{2} &=& \left.\frac{M \cos^2\theta}{fH_{1}} \right[ (a_{2}
\cosh\delta_{1}\sinh\delta_{5}\cosh\delta_{p} -
a_{1}\sinh\delta_{1}\cosh\delta_{5}\sinh\delta_{p}) dt \nonumber \\
& & +( a_{1}\sinh\delta_{1}\cosh\delta_{5}\cosh\delta_{p} -
a_{2}\cosh\delta_{1}\sinh\delta_{5}\sinh\delta_{p})dy \mbox{\huge
]}\wedge d\psi \nonumber \\
&+& \left.\frac{M\sin^2\theta }{f H_{1}} \right[ (a_{1}
\cosh\delta_{1}\sinh\delta_{5}\cosh\delta_{p} -
a_{2}\sinh\delta_{1}\cosh\delta_{5}\sinh\delta_{p}) dt \nonumber \\
& & +( a_{2}\sinh\delta_{1}\cosh\delta_{5}\cosh\delta_{p} -
a_{1}\cosh\delta_{1}\sinh\delta_{5}\sinh\delta_{p})dy \mbox{\huge
]}\wedge  d\phi \nonumber \\
  &-& \frac{M \sinh2\delta_{1} }{2fH_{1}}  dt\wedge dy - \frac{M
\sinh
2\delta_{5}}{2f H_{1}}\left(r^2+a_{2}^2+M\sinh^2\delta_{1}\right)\cos^2\theta d\psi\wedge d\phi \nonumber \\
e^{2\Phi} &=& \frac{H_{1}}{H_{5} } \label{ne} 
\eea
Here
\be
f=r^2+a_1^2\sin^2\theta+a_2^2\cos^2\theta
\ee
\be
H_i\equiv 1+K_i = 1+{M\sinh^2\delta_i\over f}, ~~~i=1,5
\ee
\section{The Extremal Limit} \setcounter{equation}{0}

\subsection{Conserved charges}

We want to take an extremal limit of the above solution. We take this limit while keeping  the conserved charges fixed to the values that describe the states $|n\rangle^{total}$ defined in  section 2. Thus the solution should describe $n_1$ D1 branes, $n_5$ D5 branes, $n_p=n(n+1)n_1n_5$ units of momentum, and angular momenta
\be 
J_\psi=-{\bar j}+j=-n~n_1n_5, ~~~J_\phi=-{\bar j}-j=(n+1)~n_1n_5
\ee
 The volume of the $T^4$ is $V$ and the length of the $S^1$ is $2\pi R$. The 10-D Newton's constant is $G^{(10)}=8\pi^6 g^2\alpha'^4$. If we dimensionally reduce along $T^4, S^1$ then we get the 5-D Newton's constant
 \be
 G^{(5)}={G^{(10)}\over V (2\pi R)}
 \ee
From the solution given above we get
\bea
Q_1&\equiv &{M\over 2}\sinh 2\delta_1= {(2\pi)^4 g \alpha'^3\over V} n_1\\
Q_5&\equiv& {M\over 2}\sinh 2\delta_p=g\alpha' n_5\\
Q_p&\equiv&{M\over 2}\sinh 2\delta_p={(2\pi)^4g^2\alpha'^4\over VR^2}n_p={(2\pi)^4g^2\alpha'^4\over VR^2}n(n+1)n_1n_5
\eea
It will be useful to define the length scale
\be
a={\sqrt{Q_1Q_5}\over R}
\ee
We observe that
\be
Q_p=n(n+1) a^2
\ee
The angular momenta are
\bea
J_\psi&=&-M \bigl(a_{1} \cosh\delta_{1} \cosh\delta_{5} \cosh\delta_{p} -
a_{2} \sinh\delta_{1}\sinh\delta_{5}\sinh\delta_{p} \bigr){\pi\over 4 G^{(5)}}=-n\,n_1n_5\\
J_\phi&=&-M\bigl(a_{2} \cosh\delta_{1} \cosh\delta_{5} \cosh\delta_{p} -
a_{1} \sinh\delta_{1}\sinh\delta_{5}\sinh\delta_{p} \bigr){\pi\over 4 G^{(5)}}=(n+1)\,n_1n_5\nonumber 
\label{ang}
\eea
It will be helpful to define
\bea
&&\!\!\!\!\!\!-{M\over\sqrt{Q_{1}Q_{5}}}\,\bigl(a_{1} \cosh\delta_{1} \cosh\delta_{5} \cosh\delta_{p} -
a_{2} \sinh\delta_{1}\sinh\delta_{5}\sinh\delta_{p} \bigr) \equiv \gamma_1 \nonumber \\
&&\!\!\!\!\!\!-{M\over\sqrt{Q_{1}Q_{5}}} \bigl(a_{2} \cosh\delta_{1} \cosh\delta_{5} \cosh\delta_{p} -
a_{1} \sinh\delta_{1}\sinh\delta_{5}\sinh\delta_{p} \bigr) \equiv \gamma_2~~~~~~
\label{alimit}
\eea
Then (\ref{ang}) implies
\be
 \gamma_1 = -a\,n\,,\quad \gamma_2=a\,(n+1)
\label{values}
\ee

\subsection{Taking the extremal limit}

To get the extremal limit we must take
\be 
M\rightarrow 0, ~~~\delta_i\rightarrow\infty~~~~(i=1,5,p)
\ee
keeping the $Q_i$ fixed.  This gives
\bea
\cosh^2\delta_i&=&{Q_i\over M}+{1\over2}+O(M)\nonumber\\
\sinh^2\delta_i&=&{Q_i\over M}-{1\over2}+O(M)
\label{deltaexpansion}
\eea
We must also take suitable limits of $a_1, a_2$ so that the angular momenta are held fixed. It is useful to invert (\ref{alimit}):
\bea
&&a_{1} = -\frac{\sqrt{Q_{1}Q_{5}}}{M} \frac{\gamma_1\cosh\delta_{1} \cosh\delta_{5}\cosh\delta_{p} +
\gamma_2 \sinh\delta_{1}\sinh\delta_{5}\sinh\delta_{p}}{\cosh^2\delta_{1}\cosh^ 2\delta_{5}\cosh^2\delta_{p} -
\sinh^2\delta_{1}\sinh^2\delta_{5}\sinh^2\delta_{p} }\nonumber\\
&& a_{2} = -\frac{\sqrt{Q_{1}Q_{5}}}{M} \frac{\gamma_2 \cosh\delta_{1} \cosh\delta_{5}\cosh\delta_{p} +
\gamma_1 \sinh\delta_{1}\sinh\delta_{5}\sinh\delta_{p} }{\cosh^2\delta_{1}\cosh^2\delta_{5}\cosh^2\delta_{p} -
\sinh^2\delta_{1}\sinh^2\delta_{5}\sinh^2\delta_{p}}~~~~~
\eea
Using (\ref{deltaexpansion}) we find
\bea
&&a_1 =-(\gamma_1+\gamma_2)\,\eta\,\sqrt{Q_p\over M} - {\gamma_1-\gamma_2 \over 4}\,\sqrt{M\over Q_p}  +
O(M^{3/2})\nonumber\\
&&\quad\,\, = - a\,\eta\,\sqrt{Q_p\over M} + a\,{2n+1\over 4}\,\sqrt{M\over Q_p}  + O(M^{3/2})\nonumber\\
&&a_2 =-(\gamma_1+\gamma_2)\,\eta\,\sqrt{Q_p\over M} + {\gamma_1-\gamma_2 \over 4}\,\sqrt{M\over Q_p}  +
O(M^{3/2})\nonumber\\
&&\quad\,\, = - a\,\eta\,\sqrt{Q_p\over M} - a\,{2n+1\over 4}\,\sqrt{M\over Q_p} + O(M^{3/2})
\label{aexpansion}
\eea
where we have defined the dimensionless combination
\be
\eta\equiv {Q_1 Q_5\over Q_1 Q_5 + Q_1 Q_p + Q_5 Q_p}
\ee
and in the second equalities we have used the specific values for $\gamma_1$ and $\gamma_2$ given in
(\ref{values}).

We thus see that for generic values of $\gamma_1$, $\gamma_2$ and $Q_p$ the parameters $a_1$ and $a_2$
diverge when $M\to 0$.  There are two exceptions: 

\noindent (a) $Q_p=0$, which is the case considered in \cite{bal, mm}; in this case $a_1$ and $a_2$ go to finite values when $M\to 0$.

\noindent (b)  ${\bar j}\sim (\gamma_1+\gamma_2)/2=0$; in this
case $a_1$ and $a_2$ go to zero as $\sqrt{M}$ in the extremal limit. This  case was studied in \cite{cy, bmpv}.

 In the case of interest to us both $Q_p$ and $\gamma_1+\gamma_2$ are non-zero and thus
 $a_1$ and $a_2$ have divergent limits when $M\to 0$. This might seem
to pose a problem for the finiteness of the metric in the
extremal limit. Note however that $a_1$ and $a_2$ have the same divergent part, so  one of the expressions occurring in the metric
\be
a_1^2-a_2^2\rightarrow -a^2\,\eta\,(2n+1)
\ee
is seen to be finite.
Note that we need to keep terms up to $O(\sqrt{M})$ in the expansion of $a_1, a_2$ in obtaining the above limit; higher order terms can however be discarded. Similar care has to be taken in other computations below.

We also encounter the expressions $r^2 + a_i^2$. Define a new radial coordinate
\be
 r^2_N = r^2 + {a^2 \eta^2 Q_p\over M}-\beta^2
\ee
The term which
diverges when $M$ goes to zero has been chosen to cancel the divergence of $a_1^2$ and $a_2^2$, and $\beta$ is a finite constant that we will fix later. Note that $r^2$ goes to $-\infty$ down the throat in the extremal limit but $r_N^2$ will reach a finite value. We find the following as  $M\to 0$:
\bea
r^2+a_1^2&=&r^2_N + \beta^2 - a^2 \eta {2n+1\over 2}\nonumber\\
r^2+a_2^2&=&r^2_N +\beta^2 + a^2 \eta {2n+1\over 2} \nonumber\\
f&=& r^2 + a_{1}^2 \sin^2\theta + a_{2}^2 \cos^2\theta \nonumber\\
\quad &=&  r^2_N + \Bigl(\beta^2 - a^2 \eta {2n+1\over 2}\Bigr) \sin^2\theta +
\Bigl(\beta^2 + a^2 \eta {2n+1\over 2}\Bigr) \cos^2\theta~~~~
\eea

The coefficient of $dr^2$ has the factor ${r^2\over (r^2+a_1^2)(r^2+a_2^2)-Mr^2}$. If we choose
\be
\beta^2={\eta\,a^2\over 2}
\label{beta}
\ee 
 we are left with only a quadratic in the denominator.  To see this note that the denominator is
\bea
&&(r^2+a_1^2)(r^2+a_2^2)-M r^2\nonumber\\
&&\qquad =\Bigl(r^2_N + \beta^2 - a^2 \eta {2n+1\over 2}\Bigr)\Bigl(r^2_N +\beta^2 + a^2 \eta
{2n+1\over 2}\Bigr) -M(r_N^2-{a^2\eta^2Q_p\over M}+\beta^2)\nonumber\\
&&\qquad\rightarrow \Bigl(r^2_N + \beta^2 - a^2 \eta {2n+1\over 2}\Bigr)\Bigl(r^2_N +\beta^2 + a^2 \eta
{2n+1\over 2}\Bigr) +{a^2\eta^2Q_p}\nonumber\\
&&\qquad=r^2_N (r^2_N + \eta\,a^2)
\label{denominator}
\eea
The numerator is $r^2dr^2=r_N^2dr_N^2$,
and we get a cancellation of the factors $r_N^2$. We will see below that in the extremal metric the point $r_N=0$ acts like an origin of polar coordinates, so the choice (\ref{beta}) is the correct one to define a coordinate $r_N$ with range $(0, \infty)$. 

We also find that other terms in the metric and gauge field are finite in the extremal limit; this can be verified using
 (\ref{deltaexpansion}),(\ref{aexpansion}). We get the extremal solution (in the string frame)

\bea
ds^2 &=& -\frac{1}{h} (dt^2-dy^2) + \frac{Q_{p}}{h f}\left(dt-dy\right)^{2} + hf
\left( \frac{dr^2_N}{r_{N}^2 + a^2 \eta } + d\theta^2 \right)\nonumber \\
 &+& h \left( r_{N}^2 - n a^{2}\eta + \frac{(2n+1)a^2\eta Q_{1}Q_{5} \cos^2\theta}{h^2 f^2} \right)
\cos^2\theta d\psi^2   \nonumber \\
&+& h\left( r_{N}^2 + (n+1) a^2 \eta - \frac{ (2n+1)a^2\eta Q_{1}Q_{5} \sin^2\theta}{h^{2} f^{2} }\right)
\sin^2\theta d\phi^2  \nonumber \\
&+& \frac{a^{2} \eta^2 Q_p}{h f} \left( \cos^2\theta d\psi + \sin^2\theta d\phi \right)^{2}  \nonumber\\
&+& \frac{2 a \sqrt{Q_{1}Q_{5}} }{hf} \left[ n \cos^2\theta d\psi - (n+1) \sin^2\theta d\phi\right] (dt-dy) 
\nonumber \\
&-& \frac{2 a \eta \sqrt{Q_{1}Q_{5}}}{h f} \left[ \cos^2\theta d\psi + \sin^2\theta d\phi \right] dy +
\sqrt{\frac{H_{1}}{H_{5}}} \sum_{i=1}^{4} dz_{i}^2 
\label{em}\\
C_{2} &=& \frac{a \sqrt{Q_{1} Q_{5}} \cos^2\theta}{H_1 f } \left( -(n+1) dt + n dy \right) \wedge d\psi 
\nonumber\\
&+& \frac{a \sqrt{Q_{1}Q_{5} } \sin^2\theta }{H_1 f } \left( n  dt - (n+1) dy \right) \wedge d\phi 
\nonumber \\
&+& \frac{a \eta Q_{p} }{\sqrt{Q_{1}Q_{5}}H_1 f } \left( Q_{1} dt + Q_{5} dy \right) \wedge
\left( \cos^2\theta d\psi  + \sin^2\theta d\phi \right)  \nonumber \\
&-& \frac{ Q_{1} }{H_1 f} dt \wedge dy - \frac{ Q_{5}\cos^2\theta }{H_1 f} \left( r_{N}^{2} + (n+1) a^2 \eta +
Q_{1} \right)  d\psi \wedge d\phi \label{extremalramond}\\
e^{2\Phi} &=& \frac{H_{1}}{H_{5}}
\label{dilaton}
\eea
\bea
f &=& r_{N}^2 - a^2\eta \,n \sin^2\theta + a^2 \eta\, (n+1) \cos^2\theta \nonumber\\
h &=& \sqrt{H_{1} H_{5}}, \ H_{1} = 1+ \frac{Q_{1}}{f}, \ H_{5} = 1+ \frac{Q_{5}}{f}
\label{deffh}
\eea

\section{Regularity of the Solution}\setcounter{equation}{0}

\subsection{Regularity of the metric}

A sufficient condition for the metric to be regular is that the coefficients of both the metric and the inverse
metric be twice differentiable functions of the coordinates. In turn, the inverse metric is well-defined if the metric is
smooth and the determinant is non-vanishing. For the metric (\ref{em})
\be
\sqrt{-g}={H_1\over H_5}hfr_N \sin\theta\cos\theta 
\label{det}
\ee

The function $f$ does vanish on a hypersurface, but the combination $hf$ takes a finite value when $f\rightarrow 0$
\be
f\to 0 \Rightarrow h f\to \sqrt{Q_1 Q_5}
\ee
The function $f$ appears explicitly in the metric only in the combination $hf$; thus we do not get singularities 
at $f\rightarrow 0$. 

A little algebra shows that 
\be
Q_1+f>0, ~~~Q_5+f>0
\label{pos}
\ee
everywhere. (To check this one needs to note that $n$ is an integer; for fractional $n$ the above expressions can become negative
at some points.) Thus the factor ${H_1\over H_5}$ in (\ref{det}) is regular and nowhere vanishing.

 We have factors of $h$ in the second and third lines of (\ref{em}).
Since
\be
h={\sqrt{(Q_1+f)(Q_5+f)}\over f}
\ee
 we have $h\rightarrow \infty$ as $f\rightarrow 0$. But a little algebra shows that when $f\rightarrow 0$ the brackets multiplying $h$ reduce to $f$ in each case, so we again get the finite combination $hf$.

The vanishing of $\sqrt{-g}$ for $\theta=0$ and $\pi/2$ does not correspond to
a singularity of the metric but only signals the degeneration of polar
coordinates at the north and south pole of $S^3$.

The most nontrivial limit arises at  $r_N \to 0$.
Let
\be
\phi\to{\tilde \phi}=\phi+{a\,n \over \sqrt{Q_1 Q_5}}\,y\,\quad \psi\to
{\tilde \psi} =\psi-{a\,(n+1)\over \sqrt{Q_1 Q_5}}\,y
\label{angularchange}
\ee
In these coordinates, the metric expanded around $r_N\to 0$ has the following
form:
\bea
&&\!\!\!\!\!\!\!\!\!\!ds^2 =
{h f\over \eta a^2}\,\Bigl(dr_N^2 + {a^2\over Q_1 Q_5}\,r_N^2\,dy^2\Bigr) \nonumber \\
&& \quad + h f\,\Bigl(d\theta^2 + {\tilde g}_{\psi\psi}\,d{\tilde \psi}^2 + {\tilde g}_{\phi\phi}\,
d{\tilde \phi}^2 + 2 {\tilde g}_{\psi\phi}\,d{\tilde \psi}\,d{\tilde \phi}\Bigr)\nonumber\\
&&\quad + g_{tt}\,dt^2  + 2 g_{t\psi}\,dt\,d{\tilde \psi} + 2 g_{t\phi}\,dt\,d{\tilde \phi}+ds^2_{T^4}
\label{r=0metric}
\eea
where now
\be
f=  \eta a^2 (n+1) \,\cos^2\theta - \eta a^2 \,n \sin^2\theta
\ee
The coefficients ${\tilde g}$ and $g$ in the equation above are differentiable
functions of $r_N$ and $\theta$ that do not vanish as $r_N\to 0$ for generic values
of $\theta$; their explicit form is not important for our argument.

There are several points to note about the above expansion:

\medskip

(a) Before the coordinate transformation (\ref{angularchange}) we had the identifications
$(y,\psi, \phi)\sim (y+2\pi m_3 R, \psi+2\pi m_1, \phi+2\pi m_2)$. Recalling that $R={\sqrt{Q_1Q_5}\over a}$ we see that
the transformation (\ref{angularchange}) gives new variables that  have a similar identification
\be
(y,\tilde\psi,\tilde \phi)\sim ( y+2\pi m'_3 R, \tilde\psi+2\pi m'_1, \tilde\phi+2\pi m'_2)
\ee

(b) Noting that $y$ is periodic with period $2\pi R$ and recalling that $R={\sqrt{Q_1Q_5}\over a}$ we
see that the coefficient of $dy^2$ is just right to make $r_N,y$ polar coordinates at the origin in the $r_N,y$ space
\be
{h f\over \eta a^2}(dr_N^2 + {a^2\over Q_1 Q_5}\,r_N^2\,dy^2) =  {h f\over \eta a^2}(dr_N^2 + {r_N^2\over R^2}\,dy^2)
\ee
and thus there is no conical defect singularity at $r_N=0$.

(c) There is no term $dtdy$ at leading order in $r_N$; this cross term is absorbed entirely in the terms
$d\tilde\psi dy, d\tilde\phi dy$. Note that if we did have a residual term $dtdy$ we would not get smoothness where the $y$ 
circle shrinks. Indeed, to reach local orthogonal coordinates we would need to define new coordinates
\be
y'=y, ~~~t'=t+cy, ~~~c\ne 0
\label{tdef}
\ee
 The identification vector in the $(y,t)$ space is $(y,t)\sim (y+{2\pi  R}, t)$, which implies the identification
\be
(y',t')\sim (y'+{2\pi  R}, t'+{2\pi c  R})
\label{disclin}
\ee
Such an identification gives a singularity at $r_N=0$ where the $y$ circle shrinks, and also generates
closed timelike curves if $t'$ is a timelike direction. Since we get $c=0$, we avoid both of these potential problems.

\medskip

Nothing special happens at $r_N=0,~ \theta=0, \pi/2$; we just have the degeneration of spherical coordinates at the poles.
We conclude that the metric is regular everywhere.

\subsection{Absence of closed timelike curves}

In the past, geometries have been constructed with D1-D5-P charges and angular momentum, and it was found that
for high enough angular momentum the geometries had closed timelike curves \cite{her, dyson, emparan1}. Our geometries on the other hand are conjectured to be dual to actual microstates of the CFT, so we expect them to be free of pathologies.

There are three periodic variables in our geometry -- $y, \psi, \phi$. Following the path
\be
(\delta y, \delta \psi, \delta \phi)=({n_1 R}\epsilon, n_2\epsilon, n_3\epsilon)
\label{vector}
\ee
gives a closed curve for any integers $n_i$. This curve is spacelike at large $r_N$, but a curve like this gave a closed timelike curve at small $r_N$ in the metric studied in \cite{dyson}. We thus ask if the vector (\ref{vector}) can become timelike anywhere;
if it does become timelike for some $r_N, \theta$ then we get a closed timelike curve. If the vector (\ref{vector}) becomes timelike for nonintegral $n_i$ then we get a timelike path that passes arbitrarily close to itself, which is pathological too; thus we investigate if the vector
\be
V^y=\alpha, ~~V^\psi=\beta, ~~~V^\phi=\gamma
\label{vectorp}
\ee
can become timelike anywhere. Note that adding components $V^r, V^\theta$  makes $V$ more {\it spacelike}, while having a part $V^t\ne 0$ makes the path non-closed. 

The norms of the vectors (\ref{vectorp}) are given by the metric $g_{ab}$ restricted to $a,b=y,\psi, \phi$. 
Let $\tilde g$ be the determinant of the metric restricted to these three  coordinates. 
The  components $g_{ab}$ of the restricted metric are regular and $\tilde g$ is positive at infinity. If some direction in the 3-torus spanned by $y, \psi, \phi$ is to become timelike then we must have a vanishing of $\tilde g$ somewhere. But an explicit evaluation of this determinant gives
\be
{\tilde g} = {r_N^2\sin^2\theta\cos^2\theta\over \sqrt{(Q_1+f)(Q_5+f)}}\,\Bigl[(r_N^2+\eta a^2)
(f+Q_1+Q_5+a^2\,n (n+1))+{Q_1 Q_5\over \eta}\Bigr]
\ee
The fact that the combinations $Q_1+f$ and $Q_5+f$ are always positive ensures that the determinant above only 
vanishes at $r_N=0$ or $\theta=0$ or $\pi/2$. Our explicit analysis of the metric in the regions around $r_N=0$ and
$\theta=0$ or $\pi/2$ shows that no closed time-like curves appear even at these points; the vanishing of $\tilde g$
just signals the vanishing of one of the three periodic coordinates.

\subsection{Regularity of $\Phi, C_{AB}$}

The dilaton  given by (\ref{dilaton}) can be written as
\be
e^{2\Phi}={Q_1+f\over Q_5+f}
\ee
From (\ref{pos}) we see that $\Phi$ is regular everywhere.

Let us finally check that the RR field is not singular. Since its
coefficients are differentiable functions of $r_N$ and $\theta$ for all values of
the coordinates, singularities can come only from places where the coordinate
system degenerates, i.e. at $\theta=0$ and $\pi/2$ (nothing special happens at $r_N=0$).

At $\theta=\pi/2$ we see that the components of $C$ are regular. 
At $\theta=0$, where the $\phi$ coordinate becomes ill
defined, the only component that is potentially singular is $C_{\phi \psi}$:
\bea
&&C_{\phi \psi}=\frac{Q_5}{f + Q_1}\Bigl(r_N^2+ (n+1) \eta a^2+ Q_1 \Bigr)\cos^2\theta  \nonumber\\
&&\qquad = Q_5 \cos^2\theta + {Q_5\,a^2\over f+Q_1}\,(1+2n)\,\sin^2\theta\,\cos^2\theta\nonumber\\
&&\qquad \approx Q_5 (1-\theta^2)+ {Q_5\,a^2\over f+Q_1}\,(1+2n)\,\theta^2\,=Q_5+O(\theta^2)
\eea
The leading term can be removed by  a gauge transformation:
\be
C_{\phi \psi}\to C_{\phi \psi} - Q_5
\ee
and $C_{\phi\psi}$ becomes regular. This is just the gauge transformation arising from a magnetic monopole potential.
Such a potential is expected since the D5 branes produce a magnetic charge for $C_2$.

\subsection{Absence of a horizon}

Since the geometry is regular, singularity theorems suggest that there is no horizon. We can check  the absence of a horizon explicitly in the following way. There is no horizon if from every point in the geometry we can find a path to asymptotic infinity
such that the tangent along the path lies in the forward light cone everywhere. 

To show that there  is such a path, we will argue that at each point in the spacetime we can find a timelike vector in the forward light cone which
has a nonzero positive $V^{r}$; thus following these vectors we can reach $r_N\rightarrow\infty$ along a timelike path.
Suppose we can find a continuous timelike vector field $V$ with $V^{r}=0$. Since $dr_N$ appears in the metric only in the form ${hfdr_N^2\over r_N^2+a^2\eta}$ we see that we can add a part $V^{r}>0$ to the vector at each point  while still keeping it timelike. (If the vector field is continuous then it lies in the \emph{forward} light cone everywhere if we choose it to lie in the forward light cone at infinity.)

We will find the  vector field $V$ in the subspace spanned by the coordinates $t, y, \psi, \phi$. 
The determinant of the metric restricted to these four directions is
\be
\hat{g}= -r_N^2\,(r_N^2+\eta\,a^2)\,\sin^2\theta\,\cos^2\theta
\ee
and is seen to be negative everywhere. Thus the tangent plane spanned by these directions has signature
$(-1, 1,1,1)$ everywhere. The region of interest is spanned by a single coordinate patch
(apart from coordinate degeneration at the poles; this causes no difficulties for the argument). We  can thus find a timelike direction in each tangent plane, which is continuous and in the forward light cone at $r_N\rightarrow \infty$. 

As mentioned above, adding a sufficiently small $V^{r}$ keeps this vector field timelike and in the forward cone
(it is important that the coefficient of $dr_N^2$ does not diverge anywhere, so the allowed $V^{r}$ can be bounded below).
Following this vector field we reach $r_N\rightarrow \infty$, and thus there is no horizon.

\section{Comparison with the solution in \cite{lunin}}

The metric in \cite{lunin} was expressed somewhat implicitly, but with some algebraic manipulation it can be brought to the form
\bea
&&\!\!\!\!\!\!\!\!\!\!ds^2 = -\frac{dt^2-dy^2}{h}+\frac{\nu(\nu+1)\,{\tilde a}^2}{h f}\,(dt+dy)^2 + 
h f \Bigl(\frac{d r^2} {r^2 + {\tilde a}^2} +  d\theta^2\Bigr) \nonumber \\
&& \quad +h \,\Bigl[ r^2 - \nu\,{\tilde a}^2+
\frac{(2\nu+1)\,{\tilde a}^2\,Q^2 \cos^2\theta}{h^2 f^2} \Bigr] \cos^2\theta d\psi^2  \nonumber \\
&& \quad + h\,\Bigl[ r^2 + (\nu+1)\,{\tilde a}^2 - 
 \frac{(2\nu+1)\,{\tilde a}^2\,Q^2 \sin^2\theta}{h^2 f^2}  \Bigr] \sin^2\theta d\phi^2  \nonumber \\
&& \quad + \frac{\nu(\nu+1)\,{\tilde a}^4}{h f} (\cos^2\theta d\psi - \sin^2\theta d\phi)^{2}  \nonumber\\
&& \quad - \frac{2\,{\tilde a}\, Q \cos^2\theta}{h f} 
\Bigr[\nu\,{Q-(1+\nu)\,{\tilde a}^2\over Q} dt  + (\nu+1)\,{Q-\nu\,{\tilde a}^2 \over Q }dy \Bigr]d\psi  \nonumber \\
&& \quad - \frac{2\,{\tilde a}\, Q \sin^2\theta}{h f}\Bigl[(\nu+1)\,{Q+\nu\,{\tilde a}^2\over Q} dt + \nu\,{Q+(\nu+1)
{\tilde a}^2\over Q} dy \Bigr]d\phi 
\label{lm}
\eea
where
\bea
&&f=r^2 + {\tilde a}^2\,(\nu+1)\cos^2\theta - {\tilde a}^2 \,\nu\sin^2\theta\nonumber\\
&&h=1+{Q\over f}
\eea
We have  denoted by ${\tilde a}$ what in \cite{lunin} has been called  $a$, to distinguish it from the analogous parameter 
appearing in our metric (\ref{em}). {\it Locally}, the metric above  reduces to ours (if we set $Q_1=Q_5=Q$ in our metric)
 after the
following redefinitions of parameters and change of coordinates:
\bea
&& {\tilde a}^2 = a^2\,\eta\,,\quad \nu = n\,,\quad \psi\to-\psi, \quad r=r_{N} \nonumber\\
&&t\to t\,\cosh\,\delta - y\,\sinh\,\delta\,,\quad y\to -y\,\cosh\,\delta + t\,\sinh\,\delta\,
\label{lunin2us}
\eea 
with
\be
e^{-2\delta}=\eta
\ee
So essentially the two metrics are related by a boost in the $y$ direction. But since $y$ is a compact direction this boost is not a symmetry; identification of points is altered by the boost.  It has been argued in \cite{lunin} that the solution is smooth; it is not immediately obvious to us though how the  identifications used there avoid shifts like (\ref{disclin}) and the corresponding potential singularities.  The charges
of \cite{lunin} appear to differ from the ones that we have, so the solutions there might correspond to a somewhat different class of 3-charge states.

\section{Discussion}\setcounter{equation}{0}

We have constructed extremal D1-D5 -P solutions dual to a special subset of CFT states. The solutions were
smooth, with no singularity or horizon.  The solutions  thus look like Fig.1(d) rather than Fig.1(c). These results support the general conjecture \cite{mss} that individual microstates
of a black hole do not look like the naive picture of a black hole -- `empty space inside the horizon with a central singularity'. 
Rather, the horizon arises only as an effective construct when we coarse-graining over microstates. Some other interesting results obtained recently also throw light on the nontrivial size associated to high entropy states \cite{bk, bena, marolf, emparan2}, in many cases by representing them as supertubes \cite{supertubes}. Topological field theories also offer hints of  a `blow up' of states in the gravity description \cite{vafa}.

The states we have looked at are a small subfamily  of all the 3-charge states. They also are not  very generic states since they carry a large angular momentum. One might therefore argue that low angular momentum states might still look like
Fig.1(c). To throw some light on this issue let us recall the case of 2-charge states. The maximally rotating D1-D5 solutions were found first {\cite{bal,mm}, by taking limits of the general solutions in \cite{cy}. These solutions turned out to be `capped' instead of extending to a pointlike singularity at $r=0$. The general D1-D5 solutions were found in \cite{lm4}; the generic solution had no angular momentum but was {\it still}  capped before reaching $r=0$. The reason can be seen best in the dual FP language. In the maximally rotating solution the F string carries the momentum P by swinging in a uniform helix, and the nonzero transverse size of the helix cuts off the geometry before $r=0$. But we can let half the F string swing clockwise and the other half anticlockwise; this still carries the same momentum P but gives no net angular momentum. Note however that we still get the nonzero transverse size, and hence the `cap' to the throat. At a technical level, it was noted in \cite{lm4} that angular momentum makes its presence felt in two kinds of terms: $\sim adtd\phi$ and $\sim {dr^2\over r^2+a^2}$. For states with no net rotation the former type of term cancels out (different parts of the geometry contribute with different signs to $a$) but the latter type of term survives since $a^2$  can only get positive contributions. We thus get a `cap' before reaching $r=0$, whether we have rotation or not. We expect that a similar situation will hold for 3-charge states, and that the naive geometry will be irrelevant inside the `horizon' radius. 

Our solutions have low curvature everywhere and the dilaton is bounded;  thus they are well described by their classical geometry. Note however that we do not expect the generic 3-charge state to be well described by a classical geometry; quantum corrections can be large. Also, while our solutions were smooth, we do not know if this will be the case for all states -- the only relevant property we seek for the generic state is that the naive geometry be not valid inside a radius of the order of the naive horizon size. The generic 2-charge FP solution was singular at the location of the string, while its dual D1-D5 system was smooth for generic configurations. The singularity at the location of the F string changed under duality to the coordinate singularity at the center of a `KK-monopole tube'  -- the geometry in the `cap' region is  a  KK monopole $\times S^1$ \cite{lmm}. Generically this $S^1$ does not self-intersect, but if we look at a special limit where the $S^1$ runs $m$ times around the same path before closing then we get the conical defect singularity arising from the coincidence of $m$ KK monopoles \cite{bal, mm, lmm}.  For the 3-charge solutions constructed in this paper we can perform dualities to interchange the D1 and P charges: $n_1\leftrightarrow n_p$. Now we have $n_p\ne n(n+1)n_1n_5 $, so we do not get the smooth solutions constructed above; we get instead extremal limits of the general solution (\ref{ne}) which have conical defects
just like the conical defects in \cite{bal, mm}.  We hope to return to an analysis of more general 3-charge solutions elsewhere.

\section*{Acknowledgments}

S.G. was supported by  an I.N.F.N. fellowship. The work of S.D.M  was supported in part by DOE grant DE-FG02-91ER-40690. We
thank Oleg Lunin and Yogesh Srivastava for helpful discussions.

\section*{Appendix A: The Non-Extremal Solution}
\begin{appendix}
\renewcommand{\theequation}{A.\arabic{equation}}
\setcounter{equation}{0}

We start with a neutral rotating black hole in 5-D lifted to 10-D
\bea
ds^2 &=& - \left(1- \frac{ M }{ f} \right) dt^2 + \frac{ r^2 f dr^2
}{(r^2+a_{1}^2)(r^2+a_{2}^2)-M r^2}   + f d\theta^2 \nonumber\\
& + & \left[ r^2+ a_{1}^2\left(1 + \frac{M}{f} \cos^2\theta\right)
\right] \cos^2\theta d\psi^2 + \frac{2M a_{1} a_{2} }{f}
\sin^2\theta\cos^2\theta d\psi d\phi \nonumber \\
& + & \left[ r^2+ a_{2}^2\left(1 + \frac{M}{f} \sin^2\theta\right)
\right] \sin^2\theta d\phi^2 + \frac{ 2 M}{f}dt\left( a_{1}
\cos^2\theta d\psi + a_{2} \sin^2\theta d\phi \right) \nonumber \\
&+& dy^2 + \sum_{i=1}^{4}dz_{i}^2
\eea
where
\be
f= r^2+ a_{1}^2 \sin^2\theta + a_{2}^2 \cos^2\theta
\ee
The parameters $a_{1},a_{2}$ give the two angular momenta. The dilaton
and the gauge fields vanish. We generate
 momentum along  $y$ by boosting
\bea
t &=& t' \cosh \delta_{5} - y'\sinh \delta_{5}  \nonumber \\
y  &=&  -t'\sinh \delta_{5} + y'\cosh \delta_{5}
\eea
and perform a T-duality $T_y$ along $y$. We get the geometry of D1 branes
\bea
ds_{10}^2 &=& - \left(1- \frac{M}{ f} \right)\frac{ dt^2}{1+K_{5}} +
\frac{ 2 M\cosh \delta_{5}}{f(1+K_{5})}\omega dt + 
\frac{dy^2}{1+K_{5}} \nonumber \\
&+& \frac{ r^2 f dr^2 }{(r^2+a_{1}^2)(r^2+a_{2}^2)-M r^2}   + f
d\theta^2 \nonumber\\
& + & \left[ r^2+ a_{1}^2\left(1 + \frac{M}{f(1+K_{5})}
\cos^2\theta\right) \right] \cos^2\theta d\psi^2 + \frac{M a_{1}
a_{2}\sin^2 2\theta }{2f(1+K_{5})}  d\psi d\phi \nonumber \\
& + & \left[ r^2+ a_{2}^2\left(1 + \frac{M}{f(1+K_{5})}
\sin^2\theta\right) \right] \sin^2\theta d\phi^2  + \sum_{i=1}^{4}
dz_{i}^{2}\nonumber \\
B_{2} &=& -\frac{M \sinh\delta_{5} }{f(1+K_{5}) } \left[
\cosh\delta_{5} dt + \omega \right] \wedge dy \nonumber\\
e^{2\Phi} &=& (1+K_{5})^{-1}
\eea
where
\be
\omega \equiv a_{1} \cos^2\theta d\psi + a_{2} \sin^2\theta d\phi, ~~~
K_{5} = \frac{M \sinh^2\delta_{5}}{f}
\ee

We again boost along $y$ with parameter $S^1$, then do an S-duality, followed by the T-dualities 
$T_{1234}$. We get the D5-P solution
\bea
ds_{10}^2 &=& \sqrt{1+K_{5}}\left[- \left(1-
\frac{M\cosh^2\delta_{1}}{ f} \right)\frac{ dt^2}{1+K_{5}}+
\left( 1+ \frac{M \sinh^2\delta_{1}}{f}\right)  \frac{dy^2}{1+K_{5}}
\right.\nonumber \\
  &+&  \frac{ 2 M\cosh \delta_{5}\cosh\delta_{1}}{f(1+K_{5})}\omega dt
- \frac{M\sinh2\delta_{1} }{f(1+K_{5})}dt dy - \frac{ 2 M\cosh
\delta_{5}\sinh\delta_{1}}{f(1+K_{5})}\omega dy \nonumber  \\
&+& \frac{ r^2 f dr^2 }{ (r^2+a_{1}^2)(r^2+a_{2}^2)-M r^2}   + f
d\theta^2 \nonumber\\
& + & \left[ r^2+ a_{1}^2\left(1 + \frac{M}{f(1+K_{5})}
\cos^2\theta\right) \right] \cos^2\theta d\psi^2 + \frac{M a_{1}
a_{2}\sin^2 2\theta }{2f(1+K_{5})}  d\psi d\phi  \nonumber \\
& + &\left. \left[ r^2+ a_{2}^2\left(1 + \frac{M}{f(1+K_{5})}
\sin^2\theta\right) \right] \sin^2\theta d\phi^2+
(1+K_{5})^{-1}\sum_{i=1}^{4} dz_{i}^{2} \right]\nonumber \\
C_{6} &=& - \frac{M \sinh \delta_{5} }{f(1+K_{5})} \left[
\cosh\delta_{5} dt\wedge dy + \omega \wedge (-dt\sinh \delta_{1} +
dy\cosh \delta_{1} ) \right]\wedge dz_{1}\wedge\cdots\wedge dz_{4}
\nonumber \\
e^{2\Phi} &=& 1+K_{5}
\eea

The solution above has a non-zero six-form RR field which should be
converted to a two-form field before we can perform the next
S-duality. The dual two form field $C_{2}$ is defined by the equation
\be
F_{3} = *F_{7}
\ee
where $*$ is taken with respect to the metric above (in the string
frame) and $F_{3}=d C_{2}, F_{7}=d C_{6}$ are the respective field
strengths. $F_7$ has the following non-zero
components
\bea
F_{rty1234} = \frac{ M r \sinh
2\delta_{5}}{f_{5}^2}~~~~~~~~~~~~~~~~~~~~~~ &\ \ \ \ & F_{\theta t y
1234} = \frac{M (a_{1}^2-a_{2}^2) \sinh 2\delta_{5}}{2 f_{5}^2} \sin
2\theta  \nonumber \\
F_{r\psi t 1234} = - \frac{2 M r a_{1} \sinh\delta_{5}
\sinh\delta_{1}}{f_{5}^2} \cos^2\theta &\ \ \ \ & F_{r\phi t 1234} =
-  \frac{2 M r a_{2} \sinh\delta_{5} \sinh\delta_{1}}{f_{5}^2}
\sin^2\theta \nonumber \\
F_{r\psi y1234} = \frac{2 M r a_{1} \sinh\delta_{5}
\cosh\delta_{1}}{f_{5}^2} \cos^2\theta ~~  &\ \ \ \ & F_{r\phi y1234}
= \frac{2 M  r a_{2} \sinh\delta_{5} \cosh\delta_{1}}{f_{5}^2}
\sin^2\theta
  \nonumber
\eea
\bea
F_{\theta \psi t 1234} &=& - \frac{M a_{1} \sinh \delta_{5} \sinh
\delta_{1} }{f_{5}^2} (r^2+ a_{1}^2 + M \sinh^2\delta_{5} ) \sin
2\theta \nonumber \\
F_{\theta \phi t 1234} &=&  \frac{M a_{2} \sinh \delta_{5} \sinh
\delta_{1} }{f_{5}^2} (r^2+ a_{2}^2 + M \sinh^2\delta_{5} ) \sin
2\theta \nonumber \\
F_{\theta \psi y1234} &=& - \frac{M a_{1} \sinh \delta_{5} \cosh
\delta_{1} }{f_{5}^2} (r^2+ a_{1}^2 + M \sinh^2\delta_{5} ) \sin
2\theta \nonumber \\
F_{\theta \phi y 1234} &=&  \frac{M a_{2} \sinh \delta_{5} \cosh
\delta_{1} }{f_{5}^2} (r^2+ a_{2}^2 + M \sinh^2\delta_{5} ) \sin
2\theta
\eea
where
\be
f_{5} \equiv f(1+K_{5}) = r^2+ a_{1}^2 \sin^2\theta + a_{2}^2
\cos^2\theta + M \sinh^2\delta_{5}
\ee
The determinant of the metric is
\be
\sqrt{-\mbox{Det}(G)} = \frac{ r f \sin\theta\cos\theta}{\sqrt{1+K_{5}}}
\ee
The non-zero components of the $F_3$ are found to be
\bea
F_{\theta\psi\phi} &=& \frac{ M (r^2+a_{1}^2)(r^2+a_{2}^2) \sin
2\theta }{ f^2 } \sinh \delta_{5}\cosh\delta_{5}     \nonumber \\
F_{\theta y \psi} &=& \frac{  M a_{2}(r^2+a_{1}^2)\sin 2\theta  }{f^2
} \sinh\delta_{5}\sinh\delta_{1}        \nonumber \\
F_{\theta y\phi} &=&  - \frac{ M a_{1}(r^2+a_{2}^2)\sin2\theta  }{f^2
} \sinh\delta_{5}\sinh\delta_{1}        \nonumber \\
F_{\theta t\psi} &=&  - \frac{ M a_{2}(r^2+a_{1}^2) \sin2\theta }{f^2
} \sinh\delta_{5}\cosh\delta_{1}        \nonumber \\
F_{\theta t \phi} &=&   \frac{ M a_{1}(r^2+a_{2}^2) \sin2\theta
}{f^2 } \sinh\delta_{5}\cosh\delta_{1}   \nonumber 
\eea
\bea
F_{r \psi\phi} &=&  - \frac{M r (a_{1}^2-a_{2}^2)\sin^{2}2\theta
}{2f^2}\sinh\delta_{5}\cosh\delta_{5}   \nonumber \\
F_{r y \psi} &=&   \frac{2M r a_{2}\cos^2\theta }{f^2}
\sinh\delta_{5} \sinh\delta_{1} \nonumber \\
F_{r y \phi} &=&   \frac{2M r a_{1}  \sin^2\theta}{f^2}
\sinh\delta_{5} \sinh\delta_{1}  \nonumber \\
F_{r t \psi} &=&   -\frac{2M r a_{2} \cos^2\theta}{f^2}
\sinh\delta_{5} \cosh\delta_{1}  \nonumber \\
F_{r t \phi} &=&   -\frac{2M r a_{1} \sin^2\theta }{f^2}
\sinh\delta_{5} \cosh\delta_{1}
\eea
For the above field strength we find the following gauge
field\footnote{This choice is up to the usual gauge freedom. In
particular the apparent asymmetry between $a_{1}$ and $a_{2}$ in the
component $C_{\psi\phi}$ can be cured by a different gauge choice. }
\bea
&C_{\psi \phi}& = - \frac{M (r^2+a_{2}^2) \cos^2\theta}{f}
\sinh\delta_{5}\cosh\delta_{5}  \nonumber \\
&C_{y \psi}& = - \frac{M a_{2}  \cos^2\theta }{f} \sinh
\delta_{5}\sinh\delta_{1},  \ \ C_{y \phi}= - \frac{M a_{1}
\sin^2\theta }{f} \sinh \delta_{5}\sinh\delta_{1} ~~\nonumber \\
&C_{t \psi}& = \frac{M a_{2}  \cos^2\theta }{f} \sinh
\delta_{5}\cosh\delta_{1}, \ \ \ C_{t \phi}=  \frac{M a_{1}
\sin^2\theta }{f} \sinh \delta_{5}\cosh\delta_{1}
\eea
Performing an S-duality followed by $T_y$ we get  the F1-NS5 solution
\bea
ds^2 &=& - \left( 1- \frac{M}{f} \right)(1+K_{1})^{-1} dt^2 +
(1+K_{1})^{-1} dy^2 \nonumber \\
&+& f(1+K_{5})\left[\frac{ r^2 )dr^2}{(r^2+a_{1}^2)(r^2+a_{2}^2)-M
r^2} + d\theta^2\right] \nonumber \\
  &+& \left[ (r^2+a_{1}^2) (1+K_{5}) + \frac{M a_{1}^2
\cos^2\theta}{f(1+K_{1})} + (a_{2}^2-a_{1}^2)\frac{K_{5} K_{1}
}{1+K_{1} } \cos^2\theta \right] \cos^2\theta d\psi^2 \nonumber \\
&+& \left[ (r^2+a_{2}^2) (1+K_{5}) + \frac{M
a_{2}^2\sin^2\theta}{f(1+K_{1})} + (a_{1}^2-a_{2}^2) \frac{K_{5}
K_{1} }{1+K_{1} } \sin^2\theta  \right] \sin^2\theta d\phi^2
\nonumber \\
&+& \frac{2 M a_{1} a_{2} \sin^2\theta \cos^2\theta }{f(1+K_{1})}
d\psi d\phi  + \frac{2 M a_{1}
\cosh\delta_{5}\cosh\delta_{1}}{f(1+K_{1})} \cos^2\theta d\psi dt
\nonumber \\
&+& \frac{2 M a_{2} \cosh\delta_{5} \cosh\delta_{1} }{f(1+K_{1})}
\sin^2\theta d\phi dt + \frac{ 2 M a_{2} \sinh\delta_{5}
\sinh\delta_{1} }{f(1+K_{1})} \cos^2\theta d\psi dy \nonumber \\
&+& \frac{ 2 M a_{1} \sinh\delta_{5} \sinh\delta_{1} }{f(1+K_{1})}
\sin^2\theta d\phi dy  +\sum_{i=1}^{4} dz_{i}^2  
\eea
\bea
B_{2} &=& \frac{M\cos^2\theta }{f(1+K_{1})} \left( a_{2}
\sinh\delta_{5}\cosh\delta_{1} dt +
a_{1}\cosh\delta_{5}\sinh\delta_{1}dy\right)\wedge d\psi \nonumber \\
&+& \frac{M \sin^2\theta}{f(1+K_{1})} \left( a_{1}
\sinh\delta_{5}\cosh\delta_{1} dt +
a_{2}\cosh\delta_{5}\sinh\delta_{1}dy\right)\wedge d\phi \nonumber \\
&-& \frac{M \sinh\delta_{1}\cosh\delta_{1} }{f(1+K_{1})} dt\wedge dy
- \frac{M \sinh
2\delta_{5}}{2f(1+K_{1})}\left(r^2+a_{2}^2+M\sinh^2\delta_{1}\right)\cos^2\theta d\psi\wedge d\phi \nonumber \\
e^{2\Phi} &=& \frac{1+K_{5} }{1+K_{1} }
\eea
Finally, doing an S-duality and performing a boost with parameter $\delta_p$ we get the D1-D5-P solution
(\ref{em})-(\ref{dilaton}).

\end{appendix}


\begin{thebibliography}{99}
%
\bibitem{hawking}
S.~W.~Hawking,
Commun.\ Math.\ Phys.\  {\bf 43}, 199 (1975).
%
\bibitem{emission}
S.~D.~Mathur,
Nucl.\ Phys.\ B {\bf 529}, 295 (1998)
[arXiv:hep-th/9706151].
%
\bibitem{lm4}
O.~Lunin and S.~D.~Mathur,
Nucl.\ Phys.\ B {\bf 623}, 342 (2002)
[arXiv:hep-th/0109154].
%
\bibitem{lm5}
O.~Lunin and S.~D.~Mathur,
Phys.\ Rev.\ Lett.\  {\bf 88}, 211303 (2002)
[arXiv:hep-th/0202072].
%

%
\bibitem{lmm}
O.~Lunin, J.~Maldacena and L.~Maoz,
arXiv:hep-th/0212210.
%
\bibitem{stromvafa}
A.~Strominger and C.~Vafa,
Phys.\ Lett.\ B {\bf 379}, 99 (1996)
[arXiv:hep-th/9601029].
%
\bibitem{mss}
S.~D.~Mathur, A.~Saxena and Y.~K.~Srivastava,
Nucl.\ Phys.\ B {\bf 680}, 415 (2004)
[arXiv:hep-th/0311092].
%
\bibitem{lunin}
O.~Lunin,
arXiv:hep-th/0404006.
%
\bibitem{gmr}
J.~B.~Gutowski, D.~Martelli and H.~S.~Reall,
Class.\ Quant.\ Grav.\  {\bf 20}, 5049 (2003)
[arXiv:hep-th/0306235].
\bibitem{sw}
N.~Seiberg and E.~Witten,
JHEP {\bf 9904}, 017 (1999)
[arXiv:hep-th/9903224];
F.~Larsen and E.~J.~Martinec,
JHEP {\bf 9906}, 019 (1999)
[arXiv:hep-th/9905064];
J.~de Boer,
Nucl.\ Phys.\ B {\bf 548}, 139 (1999)
[arXiv:hep-th/9806104].
%
\bibitem{lm1}
O.~Lunin and S.~D.~Mathur,
Commun.\ Math.\ Phys.\  {\bf 219}, 399 (2001)
[arXiv:hep-th/0006196].
%
\bibitem{lm2}
O.~Lunin and S.~D.~Mathur,
Commun.\ Math.\ Phys.\  {\bf 227}, 385 (2002)
[arXiv:hep-th/0103169].
%
\bibitem{spectral}
A.~Schwimmer and N.~Seiberg,
Phys.\ Lett.\ B {\bf 184}, 191 (1987).
%
\bibitem{bal}
V.~Balasubramanian, J.~de Boer, E.~Keski-Vakkuri and S.~F.~Ross,
Phys.\ Rev.\ D {\bf 64}, 064011 (2001)
[arXiv:hep-th/0011217].
%
%
\bibitem{mm}
J.~M.~Maldacena and L.~Maoz,
JHEP {\bf 0212}, 055 (2002)
[arXiv:hep-th/0012025].
%
\bibitem{cy}
M.~Cvetic and D.~Youm,
Nucl.\ Phys.\ B {\bf 476}, 118 (1996)
[arXiv:hep-th/9603100];
D.~Youm,
Phys.\ Rept.\  {\bf 316}, 1 (1999)
[arXiv:hep-th/9710046];
%
M.~Cvetic and F.~Larsen,
Nucl.\ Phys.\ B {\bf 531}, 239 (1998)
[arXiv:hep-th/9805097].
%
\bibitem{bmpv}
J.~C.~Breckenridge, R.~C.~Myers, A.~W.~Peet and C.~Vafa,
Phys.\ Lett.\ B {\bf 391}, 93 (1997)
[arXiv:hep-th/9602065].
%
\bibitem{her}
C.~A.~R.~Herdeiro,
Nucl.\ Phys.\ B {\bf 582}, 363 (2000)
[arXiv:hep-th/0003063];
C.~A.~R.~Herdeiro,
Nucl.\ Phys.\ B {\bf 665}, 189 (2003)
[arXiv:hep-th/0212002].
%
\bibitem{dyson}
L.~Dyson,
JHEP {\bf 0403}, 024 (2004)
[arXiv:hep-th/0302052].
%
\bibitem{emparan1}
H.~Elvang and R.~Emparan,
JHEP {\bf 0311}, 035 (2003)
[arXiv:hep-th/0310008].

\bibitem{bk}
I.~Bena and P.~Kraus,
arXiv:hep-th/0402144.
%
\bibitem{bena}
I.~Bena,
arXiv:hep-th/0404073.
%
\bibitem{marolf}
B.~C.~Palmer and D.~Marolf,
arXiv:hep-th/0403025.
%
\bibitem{emparan2}
R.~Emparan,
JHEP {\bf 0403}, 064 (2004)
[arXiv:hep-th/0402149].
%
\bibitem{supertubes}
D.~Mateos and P.~K.~Townsend,
Phys.\ Rev.\ Lett.\  {\bf 87}, 011602 (2001)
[arXiv:hep-th/0103030];
R.~Emparan, D.~Mateos and P.~K.~Townsend,
JHEP {\bf 0107}, 011 (2001)
[arXiv:hep-th/0106012].
%
\bibitem{vafa}
R.~Gopakumar and C.~Vafa,
Adv.\ Theor.\ Math.\ Phys.\  {\bf 3}, 1415 (1999)
[arXiv:hep-th/9811131];
%
J.~M.~Maldacena and C.~Nunez,
Phys.\ Rev.\ Lett.\  {\bf 86}, 588 (2001)
[arXiv:hep-th/0008001];
%
A.~Iqbal, N.~Nekrasov, A.~Okounkov and C.~Vafa,
arXiv:hep-th/0312022.
%
\end{thebibliography}
\end{document}